\documentclass[doublecol]{epl2}
\usepackage{epsfig,graphicx,graphics,amsmath,amssymb}
\usepackage[T1]{fontenc}
\usepackage[latin9,utf8]{inputenc}
\usepackage{xcolor}
\usepackage{amscd}
\usepackage{bm}
\usepackage{bbold}
\usepackage{psfrag}
\usepackage{mathrsfs}
\usepackage{booktabs}
\usepackage{svg}
\usepackage{mathrsfs}

\usepackage{graphicx}
\usepackage{braket}
\usepackage[colorlinks=true, allcolors=blue]{hyperref}

\usepackage{dsfont}
\usepackage{blindtext}

\def\ketvec#1{\mathinner{|{#1}\rangle\rangle}}
\def\bravec#1{\mathinner{\langle\langle{#1}|}}

\newcommand{\bla}{bla\\bla\\bla\\bla\\bla}

\title{Procrustes Tomography -- reconstructing noisy quantum channels made easy}

\author{Josey Stevens \inst{1,2}\thanks{E-mail: Josey.Stevens@jhuapl.edu} \and Reece Robertson \inst{3,1,2} \and Sebastian Deffner \inst{1,2,4}}
\shortauthor{Josey Stevens, Reece Robertson, and Sebastian Deffner}

\institute{
\inst{1} Department of Physics, University of Maryland, Baltimore County, Baltimore, MD 21250, USA\\
\inst{2} Quantum Science Institute, University of Maryland, Baltimore County, Baltimore, MD 21250, USA\\
\inst{3} Department of Computer Science and Electrical Engineering, University of Maryland, Baltimore County, Baltimore, MD 21250, USA\\
\inst{4} National Quantum Laboratory, College Park, MD 20740, USA\\
}

\date{\today}

\abstract{What is the most expensive part of quantum device characterization? Clearly, the answer is quantum process tomography. However, especially for noisy intermediate-scale quantum (NISQ) computers, a comprehensive understanding of the noisy quantum dynamics is essential in interpreting the computational output. In this work, we introduce an efficient method -- Procrustes tomography -- that outperforms established methods in a number of aspects. After a pedagogical and constructive introduction, we demonstrate the utility of the method for representative examples of noisy quantum channels. }

\begin{document}

\maketitle

The availability of cloud-provided noisy intermediate-scale quantum (NISQ) computers \cite{larose2019overview, AmazonBraketDevices2026, IonQCloudDocs2026, QuantinuumDocs2026, IQMResonance2026, RigettiQCSDocs2026, wurtz2023aquila} has enabled researchers to rapidly demonstrate new theoretical results via conveniently available \textit{experimental} capabilities.  
While this has clearly accelerated the pace of discovery and validation of quantum phenomenon \cite{deffner2017demonstration,frey2022realization,chen2022error,koh2022experimental,shen2025observation,choo2018measurement,iqbal2024non,bando2020probing}, there is a great drawback compared to the more traditional route where experiments are designed and carried out by experienced experimentalists who have detailed knowledge of the underlying experimental setup and how the capabilities and limitations map onto the theoretical predictions.
To enable the NISQ-user to perform informed quantum calculations, most NISQ computer providers publish curated device information that provides insights into native gate sets, device connectivity and topology, and characterizations of device noise.
To further alleviate the gap between the design of the (computational) experiment and the theoretical predictions, most NISQ devices are accompanied by noisy simulators that attempt to simulate the performance of the noisy device. 
Unfortunately, the noisy simulators often differ greatly from the results on the real devices \cite{robertson2025simon,robertson2026variational,irfan2025quantum,bautra2026evaluating,di2023noisy}, leaving the NISQ users with the difficult problem of understanding at what stage the experiment diverged from the expected behavior.

While there is likely no single universal driving factor to explain why full device characterization is not made available to the user base across all cloud-NISQ providers, it is worth noting that doing so may be against their competitive commercial interests since a full device characterization could provide deep insight into their relative technological strengths (and weaknesses) and provide insights into their technology stack.
Similar 'trade secrets' also exist in the cloud-computing space, where information such as device architecture, datacenter communication and connectivity, and software stack is often kept as proprietary information.
Moreover, publishing highly specific device data creates liability for companies if the current calibration fluctuates too far below the public claim.
However, unlike the classical computing domain, where withholding such information is limited to impacting computational efficiency, in the quantum domain, such information can render a computation or experiment obsolete before it is even designed.
It is also likely that full device characterization is not useful to \textit{many} users since simulating the full device characteristics for large circuits requires exponentially increasing resources except for special subclasses of circuits \cite{gottesman1998heisenberg,aaronson2004improved,anders2006fast}.
It is worth noting that this also likely accounts for the discrepancy between quantum devices and their classical noisy simulators---since a simulator that cannot be run in an achievable time is of no utility.  

In principle, the NISQ-user could perform tomography \cite{nielsen2010quantum} on the device to determine the true nature of the device's performance.
While this is experimentally costly, it allows for significantly deeper insights into the realized device performance and the impact on the experimental results.
Indeed, both state and process tomography have been the subject of much development over the preceding decades, resulting in a broad range of results from introductory textbook methods to advanced methods that take advantage of machine learning \cite{torlai2023quantum,gaikwad2024neural,ahmed2023gradient} and entanglement \cite{altepeter2003ancilla,xue2022variational,wu2025error}, or that are motivated by experimental and physical insights into the noise \cite{rudinger2021experimental,moueddene2021context,cao2024efficient,vinas2026context,rodionov2015compressed,kosut2008quantum}.
Meanwhile, the most introductory methods often focus on pedagogy or historical relevancy at the cost of efficiency and ease of implementation.  
While the most advanced of these methods provide the most efficient implementations, they also require substantial expertise and require the most upfront effort to implement, and may themselves require device access not made available to the user. 
The onus of characterizing the device performance beyond what is made publicly available by the cloud provider is placed on the NISQ-user, who, aside from the impact on understanding or predicting their experimental results, might have little interest in tomography.
In these cases, a conceptually \textit{lightweight} and \textit{extensible} methodological approach to tomography that is of sufficient (if not highest) quality may be particularly desirable to these users.

In this letter, we review three approaches to process tomography that are particularly pedagogical and suitable for these purposes.
First, we provide an alternate presentation of the canonical (textbook) process tomography that has a straightforward (if nonphysical) interpretation and may provide the reader with an alternative interpretation of the standard interpretation. 
Then, we review linear-inversion process tomography, which operates directly on measurement probabilities.
Finally, we introduce and expound on a formulation of process tomography based on the solution of a Procrustes problem in terms of reconstructed density matrices. These methods are then compared with numerical simulations to investigate the scaling performance of these methods under various experimental scenarios.
We find that the performance of our novel Procrustes tomography is competitive with, and often better than, the performance of the other methods.

\section{Essentials of process tomography} \label{Process Tomography Section}

We begin by briefly describing three methods of process tomography. 

\subsection{Canonical process tomography} The standard canonical introduction to process tomography \cite{nielsen2010quantum} often centers around determining either the $\chi$-matrix or the Choi-matrix.
With an emphasis on pedagogy, we breifly review a particularly clean description for the Choi-matrix, $\rho_{\text{Choi}}$, that can be expressed in terms of un-propagated and propagated coherences ($\ket{i}\bra{j}$) via the equation
\begin{equation}\label{Eq: Choi Matrix}
\rho_{\text{Choi},\mathcal{E}} = \sum\limits_{i,j} \ket{i}\bra{j} \otimes \mathcal{E}\left( \ket{i}\bra{j} \right).
\end{equation}
This equation seems only deceptively simple, since it implies the need to create and propagate non-physical states (if $i\ne j$, $\text{Tr}(\ket{i}\bra{j}) = 0$).
To avoid this impossibility, one calculates $\ket{i}\bra{j}$ as a linear combination of physical states, for example, from the well-known relationship \cite{nielsen2010quantum,mohseni2008quantum}
\begin{equation} \label{Eq: Coherence expansion}
\ket{n}\bra{m} = \ket{+}\bra{+} \ + \ i\ket{-}\bra{-} \ - \ \frac{1+i}{2} \left(\ket{n}\bra{n} + \ket{m}\bra{m}\right),
\end{equation}
where $\ket{+} = (\ket{n}+\ket{m})/\sqrt2$ and $\ket{-} = (\ket{n}+i\ket{m})/\sqrt2$.

Since completely positive trace-preserving (CPTP) maps are linear, i.e., $\mathcal{E}\left(\sum_i \rho_i \right) = \sum_i\mathcal{E}(\rho_i)$, one can prepare the requisite states---$\ket{n}$, $\ket{+}$, and, $\ket{-}$---propagate them through the noisy process, and then use state tomography to estimate $\mathcal{E}(\ket{n})$, $\mathcal{E}(\ket{+})$, and $\mathcal{E}(\ket{-})$, respectfully.
Once estimated, $\mathcal{E}(\ket{i}\bra{j})$ can be calculated via Eq.~\eqref{Eq: Coherence expansion}, which can be utilized to calculate $\rho_\text{Choi}$ with Eq.~\eqref{Eq: Choi Matrix}.
If the Choi-matrix is decomposed into its spectral decomposition, $\rho_{\text{Choi}} = \sum_i \lambda_i \ket{\psi_i}\bra{\psi_i}$, then the vectorized Kraus operators can be defined as $\ketvec{K_i} = \sqrt{\lambda_i}\ket{\psi_i}$, and with this a full characterization of the process is obtained.  

\subsection{Linear inversion process tomography} The second popular version of process tomography is based on the determination of measurement outcomes (with projector $E_i$) relative to some density operator $\rho_j$.
Mathematically, this is defined as
\begin{equation} \label{Eq: individual probabilities}
    p_{i,j} = \text{Tr}(E_i\rho_j') = \bravec{E_i}\ketvec{\rho_j'}.
\end{equation}
Here we use the notation such that, for an $m \times n$ operator $A$, $\ketvec{A}$ is the vectorized \cite{miszczak2011singular} form of $A$,
\begin{equation}
    \begin{aligned}
    \ketvec{A} \equiv (a_{11},&a_{21},...,a_{m1},a_{12},a_{22},\\
    &...,a_{m2},...,a_{1n},a_{2n},...,a_{mn})^T.
    \end{aligned}
\end{equation}
This allows to compute all of the pairwise measurement probabilities against all states in Eq.~\eqref{Eq: individual probabilities} with the matrix equation
\begin{equation}
    P = MA',
\end{equation}
where $M$ is a matrix whose $i$th row is given by $\bravec{E_i}$ and $A'$ is a matrix whose $j$th column is $\ketvec{\rho_j'}$. 

In this notation, passing an input state through a noisy process consists of applying the process matrix, $\Lambda$, to the input state, i.e., $\ketvec{\rho_j'} = \Lambda\ketvec{\rho_j}$.
This allows to characterize the measurement probabilities with respect only to input states (which form the columns of $A$) before they pass through the noisy process as
\begin{equation}
    P = M \Lambda A. 
\end{equation}

This equation can then be inverted to solve for $\Lambda$ as,
\begin{equation} \label{Eq: linear inversion}
    \Lambda = M^{-1}PA^{-1}.
\end{equation}
If the input states or measurements are chosen such that $M$ or $A$ are not square, one simply replaces the inverse with the pseudoinverse.
Realizing process tomography in this framework then consists of choosing a tomographically complete set of input states and measurements (choices that make Eq.~\eqref{Eq: linear inversion} a complete matrix equation) and performing enough experiments to estimate $P$ by preparing input states $\ketvec{\rho_j}$ and calculating the relative rates of measurement outcome $E_i$.

\subsection{Procrustes tomography} We now introduce \emph{Procrustes tomography},  as an efficient, and pedagogically simple form of tomography.

In Procrustes tomography, instead of working with the measurement probabilities directly, we solve for the process matrix, $\Lambda$, by working only with input and output states.
As before, we place the input states $\ketvec{\rho_j}$ into the columns of $A$ and the output sates $\ketvec{\rho_j'}$ into the columns of $B$, and then construct the propagation problem
\begin{equation}
    B = \Lambda A.
\end{equation}
Now, if one can do state tomography to estimate the states of $B$, one can solve for $A$ by solving
\begin{equation} \label{Eq: Minimization}
    \Lambda = \min_{\mathcal{L}}\left|\mathcal{L}A -B\right|_F,
\end{equation}
which is the well-studied Procrustes problem  \footnote{In Greek Mythology, Procrustes was an infamous son of Poseidon.
He is best known as a smith with a bed that he claimed could exactly fit any traveler.
Unfortunately for the travelers, Procrustes would wait until after he had strapped them to the bed to reveal his method: he would stretch the legs of anyone that was too short, or lop the legs of anyone that was too long, until the person fit the bed exactly.
The analogy to the matrix problem of Equation \eqref{Eq: Minimization} is as follows: the matrix $A$ is the proverbial traveler, the matrix $B$ is the proverbial bed, and the transformation $\mathcal{L}$ which represents the stretching/lopping required to fit $A$ onto $B$ \cite{gower_procrustes_2009}.} \cite{gower_procrustes_2009}.
In Equation \eqref{Eq: Minimization}, $|\cdot|_F$ represent the Frobenius norm.
This problem is minimized if
\begin{equation} \label{Eq: Procrustes Solution}
    \Lambda = BA^+.
\end{equation}
Like linear inversion process tomography, one prepares input states, passes them through the noisy process, and then performs state tomography on the output states to estimate $\ketvec{\rho_j'}$.

The connection between Procrustes tomography and linear inversion process tomography is not superficial---indeed, if linear inversion state tomography is utilized to estimate the output states then the mathematical construction is identical and the connection to least squares minimization has previously been recognized \cite{hashim2025practical}.  
The only exception to this fact is that formally estimating the density matrices before passing them through Eq.~\eqref{Eq: Procrustes Solution} allows one to force the compatibility of measurement probabilities with a positive-semidefinite density matrix which is not enforced if one utilizes measurement outcome statistics directly.
This is done by taking the spectral decomposition of the density matrix, zeroing out any negative eigenvalues, reconstructing the density matrix, and re-normalizing.

However, Procrustes tomography does not require the use of linear inversion state tomography; any other methods \cite{innan2024quantum,wei2024neural,lohani2020machine,schmale2022efficient} of state tomography may be utilized to estimate output states, including those that attempt to correct for challenges arising from readout errors \cite{aasen2024readout}.  
In this way, any improvement in state tomography, either in general or for a specific system, can be immediately realized as improvements in Procrustes tomography without any modification to the tomography algorithm.

It is frequently stated that standard process tomography does not account for measurement errors or state preparation errors that might arise in the preparation of the input states $\ketvec{\rho_i}$ \cite{merkel2013self,greenbaum2015introduction,huang2025quantum,hashim2025practical}.  
However, by characterizing measurement error in the state-tomography step as a systematic but unknown variance, Procrustes tomography can mitigate the impact of these errors. 
Additionally, as long as some tomographically complete set of states is prepared, Eq.~\eqref{Eq: Procrustes Solution} is robust with respect to the exact input states.
This means that accounting for state-preparation errors is as simple as performing state tomography on both the input and output states, requiring twice the number of experiments.
A similar result has recently been proposed \cite{blume2024easy} by extending linear inversion in a way that accounts for state preparation; however, in Procrustes tomography, accounting for state preparation errors requires only a trivial change to the methodology.

\subsection{Weighted Procrustes Tomography} An implicit assumption in the standard methods outlined above is that all input states or measurements are estimated to equivalent fidelity.
In the formulation of Procrustes tomography in Eq.~\eqref{Eq: Minimization} this is also the case.
However, we can introduce a weighting matrix, $W$, into Eq.~\eqref{Eq: Minimization} such that,
\begin{equation} \label{Eq: Weighted Minimization}
    \Lambda = \min_{\mathcal{L}}\left|\left(\mathcal{L}A -B\right)W\right|_F.
\end{equation}

If $W$ is a diagonal matrix with entries $w_i$, then this transforms the minimized objective function to  $\sum_iw_i^2\sum_j\left|(\mathcal{L}A-B)_{i,j}\right|^2$, against which the columns corresponding to larger $w_i$ will be optimized more aggressively.
Since we expect the standard error, $\epsilon$, of the mean of a random process to scale as $\epsilon \propto 1/\sqrt{N}$ where $N$ is the number of independent samples, if each density matrix, $\rho_i'$,  is reconstructed from $N_i$ measurement shots, we will weight each input by the fourth root of the number of samples, i.e. $w_i = \sqrt[4]{N_i}$.
If $W$ is not diagonal, then this minimization also accounts for correlations that may arise, for example, from the prepared states not being orthogonal or from correlated errors introduced in the preparation or measurement process.
This process allows us to increase the relative weighting of more well-known states---either resulting from uneven sampling in state tomography or increased variance due to measurement errors.

\subsection{Finite size sampling effects} \label{Sec: CPTP projections}

All of the methods explored so far suffer from finite-size sampling effects that result in the estimated processes not being completely positive or trace-preserving.
While these effects do fall off in the infinite-sample limit, they nonetheless are non-negligible in the intermediate-sample regime, and may also arise from non-unitary measurement errors.

In addition to their non-physical nature, non-CPTP maps can transform proper positive semi-definite density matrices into indefinite density matrices, which, when compared, can result in artificially inflated fidelity measures (in extreme cases, the fidelity can become greater than one). 
Thus, it is essential to correct these effects, even in the limit of large sample sizes, in order to evaluate the accuracy of process tomography.

Multiple approaches have been proposed to directly address these effects by either projecting processes onto the completely positive and trace-preserving sub-manifolds of superoperators \cite{PhysRevA.98.062336, Drusvyatskiy2015} or by modifying the process tomography algorithms to perform their minimization over the space of CPTP maps \cite{PhysRevA.55.R1561, PhysRevA.98.062336, j5gh-hmtw}.
In this work (for all algorithms), we utilize the algorithm provided by Knee et al. \cite{PhysRevA.98.062336} to directly project all estimated processes onto the completely positive and trace-preserving set.  

\subsection{Unitary projection}

All of these methods return a full-rank representation of a CPTP map.
In this context, we consider the rank as the number of non-zero Kraus operators needed to represent the channel.
When processes are realized in a \textit{nearly} unitary fashion, this provides a significant challenge to process tomography---for a $d$-qubit system, the maximum possible rank of the process is $d^2$.
This means that it is extremely likely that any variation due to sampling noise will be interpreted as a spurious signal best represented by a high-rank process.
If the channel is known to be \textit{nearly} unitary, the estimated process can be projected onto the nearest unitary (rank-1) process.

To demonstrate this, we utilize the process matrix estimated from Procrustes tomography via Eq.~\eqref{Eq: Procrustes Solution}.
We then want to find a solution of 
\begin{equation} \label{Eq: Rank-1 Minimization}
    \Lambda= \min_{K}\left|\Lambda -K^*\otimes K\right|_F,
\end{equation}
which is a special form of the nearest Kronecker product problem \cite{LOAN200085} and admits a solution via the Kronecker Product Singular Value Decomposition \cite{Golub1996-xu}.
If $\mathcal{R}(\Lambda)$ is the rearrangement \cite{LOAN200085} of $\Lambda$ and can be expressed via its SVD as $\mathcal{R}(\Lambda) = U \Sigma V^\dagger$, then the closest unitary can be solved as,
\begin{equation} \label{Eq: Closest Unitary}
    \Lambda_1  = \frac{\sqrt{\sigma_1}}{2}\mathcal{M}\left(U_1 + V_1^*\right),
\end{equation}
where $\mathcal{M}$ is the \textit{matrixizing} operation which reverses vectorization operation, i.e., $A = \mathcal{M}\left(\ketvec{A}\right)$, and $\sigma_1$ is the largest singular value with $U_1$ and $V_1$ being the associated vectors.
We take the average over $V_1$ and $U_1$ to ensure unitarily.

\section{Numerical simulations}\label{Sec: Numerical Simulations}

To analyze the performance of the three methods of process tomography, we perform numerical simulations of a noisy channel generated by the noisy dynamics of a driven quantum system as described by the Lindblad master equation
\begin{equation}
\dot\rho = -\frac{i}{\hbar} \left[H,\rho\right] + \sum_i\gamma_i\left(L_i \rho L_i^\dagger - \frac{1}{2}\{L_i^\dagger L_i,\rho\}\right),
\end{equation}
where $\gamma_i$ are the damping rates for the jump operators $L_i$ that describe dissipate dynamics. 
Namely we study the case of $L_1 = \sigma_- = (\sigma_x + i \sigma_y)/2$ and $L_2 = \sigma_z$ that characterize qubit amplitude damping and dephasing, respectively.
We set the associated coupling constants to $\gamma_1 = 1/T_1$ and $\gamma_2 = 1/T_2$, where $T_i$ is the associated relaxation time of the decay modes.
We determine $H$ by generating a random target unitary $U$ and setting $H$ as the most efficient constant Hamiltonian \cite{Aifer_2022, Stevens_2025} that implements the unitary in time $\tau$, i.e., $H = -(i/\hbar \tau)\text{ln}(U)$ where we use the principal branch of the matrix logarithm \cite{Higham2008-mv}.

The figure of merit we use is the average infidelity over randomly chosen input states passed through the true process ($\mathcal{E}$) and estimated porocess ($\mathcal{E'}$).
Mathematically, this is given by
\begin{equation}
1-\langle F_\mathcal{E} \rangle =1 - \int d\rho \text{F}(\mathcal{E}(\rho),\mathcal{E'}(\rho)),
\end{equation}
where $F$ is the fidelity given by $F(\sigma,\rho) = \text{Tr}\left(\sqrt{\sqrt{\rho}\sigma \sqrt{\rho}}\right)^2$.

When state-preparation errors are included, we utilize the same noise model described in the preceding discussion while also assuming that the state is prepared in time $\tau$.
Since our prepared states are either one- or two-qubit states, their generation will take (in the most general case) two-qubit gates that are of the same class as the desired gates.
In this case, we create the Hamiltonian that prepares the state with the following procedure.
A minimal non-CPTP preparation process $\Lambda^{\text{prep}}$ is calculated from the (assumed) input state $\ketvec{0}$ and the desired prepared state $\ketvec{\rho_i}$ via:
\begin{equation}
    \Lambda^{\text{prep}} = \ketvec{s}\ketvec{0}^+,
\end{equation}
where again, for an operator $A$, $A^+$ indicates the psudoinverse of $A$.
To turn this into a desired unitary process, we follow Eqs.~\eqref{Eq: Rank-1 Minimization} and \eqref{Eq: Closest Unitary} to arrive at $\Lambda^{\text{prep}}_1$.
It is worth noting that since $\ketvec{s} = \Lambda^{\text{prep}}\ketvec{0}$ is an under-constrained problem, the outputs of this procedure are not unique even though they are deterministic.
Finally we can calculate the minimal preparation Hamiltonian $H^{\text{prep}} =-(i/\hbar \tau)\text{ln}\left(\Lambda_1^{\text{prep}}\right)$.

\subsection{Results}\label{Sec: Results}

To assess the performance of the three methods, we now compare the outcome when applied to simulated experiments conducted under differing conditions.
In all cases, the infidelity is plotted as a function of number of measurement shots $N$.
Since our goal is to compare process tomography methods, we define one measurement shot as one series of state-topographically complete measurement sets with all measurements conducted in the multi-qubit Pauli basis.
In all cases, both a \textit{high}-quality (long relaxation time) and \textit{low}-quality (short relaxation time) implementation of every gate is used in order to understand the impact of gate quality on tomographic reconstruction.
For the following examples, Proctrustes tomography is performed with linear-inversion state tomography.

\begin{figure}
    \centering
    \includegraphics[width=.48\textwidth]{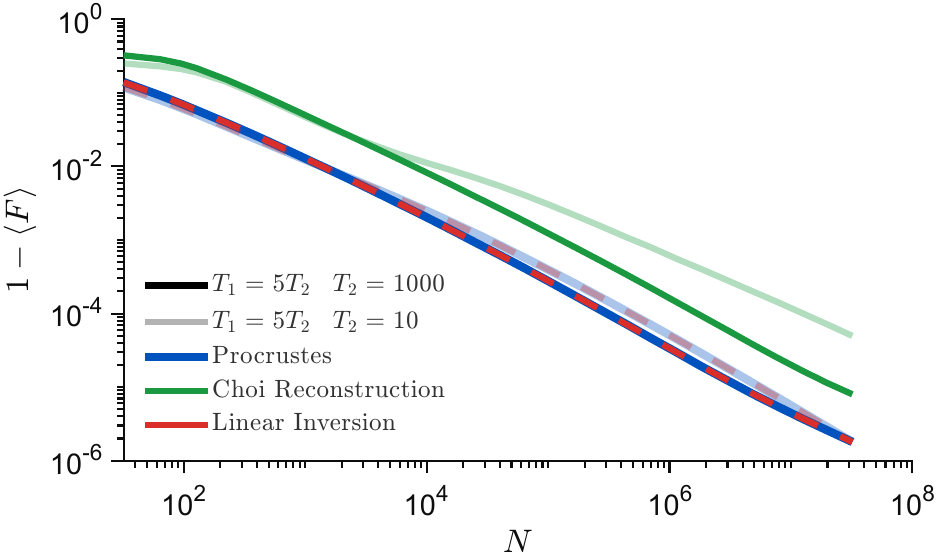}
    \caption{Infidelity between the true and reconstructed two-qubit noisy gates as a function of the number of measurement shots with evenly distributed state measurement.  Procrustes reconstruction (blue), Choi reconstruction (green), and Linear inversion (gold dashed) are plotted with high-fidelity gates for each reconstruction shown with full color saturation, and the low-fidelity implementation is plotted with reduced color saturation.
    Each method uses all 32 perfectly prepared input states required to compute the Choi reconstruction with $N_i = N_j$ for all states.}
    \label{fig: TomographyComparison1x}
\end{figure}

In Fig.~\ref{fig: TomographyComparison1x}, where states are prepared perfectly and shots are evenly sampled, we see clearly that in all cases, in the limit of large measurement shots, all methods are able to provide high-fidelity reconstructions of the noisy gates.  
In this case, to evenly compare the three methods, the full set of 32 ($\ket{i}$,$\ket{\pm}$) states required for the Choi reconstruction are also utilized for both Procrustes and Linear Inversion tomography.
In this scenario, we see that Procrustes and Linear Inversion tomography perform identically and outperform the Choi reconstruction method.
This is because even though the same states are being utilized by all methods, the Choi reconstruction utilizes them suboptimally by forming a reduced number of virtual states that then have correlated sample noise.

\begin{figure}
    \centering
    \includegraphics[width=.48\textwidth]{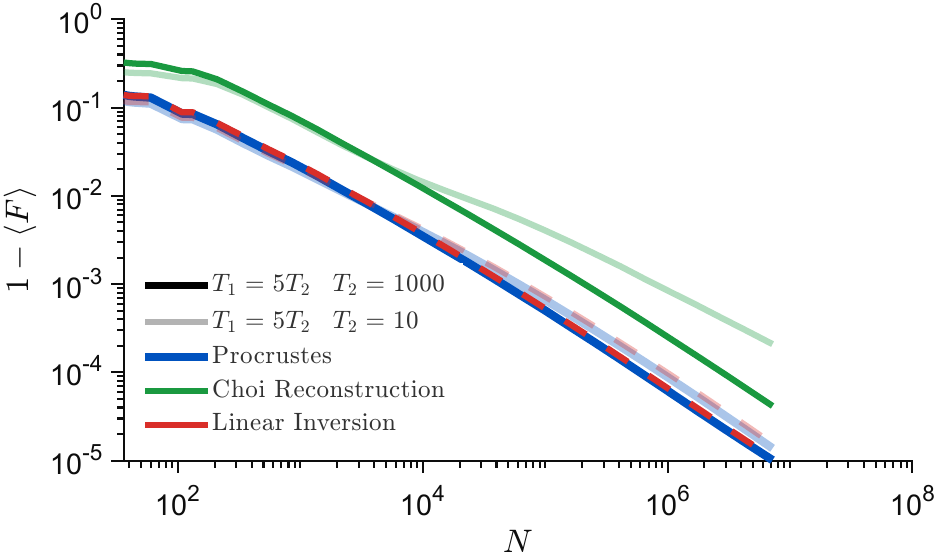}
    \caption{Infidelity between the true and reconstructed two-qubit noisy gates as a function of the number of measurement shots with unevenly distributed state measurements.  Procrustes reconstruction (blue), Choi reconstruction (green), and Linear inversion (gold dashed) are plotted with high-fidelity gates for each reconstruction shown with full color saturation, and the low-fidelity implementation is plotted with reduced color saturation.
    Each method uses all 32 perfectly prepared input states required to compute the Choi reconstruction.
    States that have a preparation of $\ket{i}{\bra{i}}$ are sampled at 10$\times$ the frequency than those representing $\ket{\pm}{\bra{\pm}}$, i.e. $N_i = 10 N_\pm$.}
    \label{fig: TomographyComparison.1x}
\end{figure}

Now consider Figu.~\ref{fig: TomographyComparison.1x}, where measurements corresponding to prepared $\ket{i}$ states are sampled at $10\times$ the number of $\ket{\pm}$---in this case we have a non-trivial weighting matrix ($w_i\ne w_j$ for all $i$,$j$).
A motivation for this kind of sampling scheme is that local states in the computational basis can be prepared more cheaply and at higher quality than the superposition states.
First, we observe that the inclusion of the weighting matrix in Procrustes tomography allows for a small but detectable systematic increase in the fidelity of the reconstructed process since the Procrustes tomography does not overreact to sampling noise from the uneven measurement shots. 
We also observe that for all methods, the infidelity of the reconstructed process is not monotonic in the number of measurement shots.
This is because at small measurement shots, the uneven sampling can provide strong over-biasing toward Clifford operations.

\begin{figure}
    \centering
    \includegraphics[width=.48\textwidth]{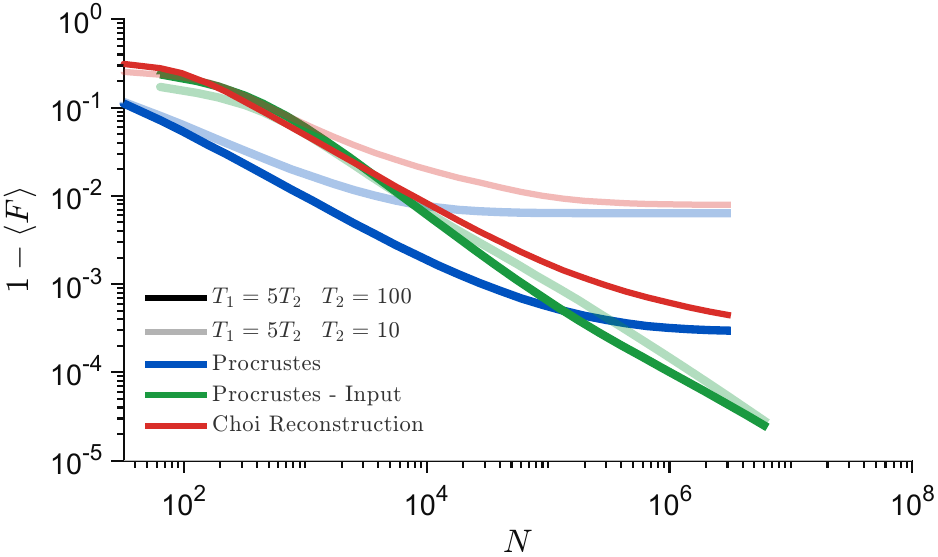}
    \caption{Infidelity between the true and reconstructed two-qubit noisy gates as a function of the number of measurement shots with evenly distributed state measurement and imperfect state preparation.  Procrustes reconstruction (blue), Choi reconstruction (green), and Linear inversion (gold dashed) are plotted with high-fidelity gates for each reconstruction shown with full color saturation, and the low-fidelity implementation is plotted with reduced color saturation.
    In each case, the state preparation is performed via an efficient unitary $U_p$ that is implemented in an identical duration to the noisy process, i.e., $\tau_P = \tau$, and is subject to the same noisy process.
    Each method uses all 32 perfectly prepared input states required to compute the Choi reconstruction with $N_i = N_j$ for all states.}
    \label{fig: Imperfect State Preparation}
\end{figure}

In Fig.~\ref{fig: Imperfect State Preparation}, we can see the impact of imperfect state preparation.
Here, Procrustes tomography is implemented in such a way that state tomography is performed with and without input state tomography.
In these scenarios, when performed without input state tomography, Procrustes tomography performs identically to Linear Inversion tomography.
We immediately see that the methods that do not account for input-state preparation errors saturate at the level of preparation error, while Procrustes tomography is able to account for these errors.
Interestingly, the cost of these measurements is not just a doubling of the number of measurement shots, but the onset of a plateau at small measurement shots where the fidelity of the reconstruction does not increase rapidly.  

\begin{figure}
    \centering
    \includegraphics[width=.48\textwidth]{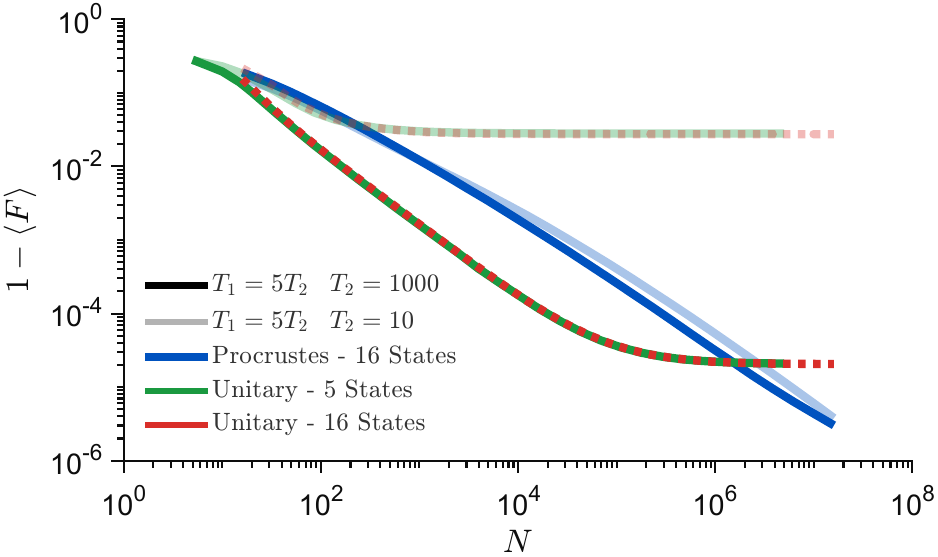}
    \caption{Infidelity between the true and the Procrustes reconstructed two-qubit noisy gates as a function of the number of measurement shots of varying numbers of input preparations including projections onto the nearest unitary operator.  
    Procrustes tomography with 16 input states is shown in blue, the unitary projected reconstruction from 5 input states ($\ket{i}$ and $H^\otimes \ket{0}$) is shown in green, while the unitary projection from the reconstruction from 16 input sates is shown in gold with high fidelity gates for each reconstruction shown with full color saturation and the low fidelity implementation of plotted with reduced color saturation and their nearest unitary projection for each case depicted as dashed lines.
    All input states are perfectly prepared with an equal distribution of measurements, i.e., $N_i = N_j$ for all states.}
    \label{fig: Near Unitary}
\end{figure}

In Fig.~\ref{fig: Near Unitary}, we depict the performance of Procrustes tomography when only a topographically complete set of states is utilized and how the performance is affected when projecting onto the unitary subspace of processes.
As expected, projected-unitary saturates infidelity at decreasing levels depending on the quality of the gate; before that point, the unitary projection significantly outperforms the full process tomography.
It has been recognized \cite{romer2025reconstructing} that a $d$-dimensional unitary process can be fully characterized by $d+1$ states---we find that utilizing this subset of states marginally outperforms the tomographically complete basis set at all measurement shot counts.

\section{Concluding remarks}

In this work, we have examined and compared the theory of three approaches to process tomography that may be of particular interest to users of NISQ-era quantum computers, including introducing a novel formulation of process tomography as a Procrustes problem in propagating reconstructed states.
We have demonstrated how this method can be modified to account for state-preparation errors and uneven sampling and demonstrate a procedure for projecting process tomography onto the nearest unitary process.
Where possible, we have provided a pedagogical exposition that relates the mathematical construction to the physical objects and processes of quantum devices.
Finally, we have performed and analyzed the process tomography methods under several experimentally motivated scenarios and constraints.

We find that Procrustes-based tomography matches or exceeds the performance of the other assessed tomography methods while having a particularly straightforward interpretation and providing a flexible framework that admits many extensions.
An additional advantage of Procrustes-based tomography is that, in certain circumstances, characterizing device state preparation and performing process tomography can be accomplished without additional effort.
Given the modern availability of noisy quantum devices that can provide rapid experimental capabilities for quantum researchers, the need for high-quality, easily interpretable, and pedagogical process tomography may enhance the ability of NISQ computers to support rapid demonstration of novel quantum phenomena.

\acknowledgments{S.D. acknowledges support from the John Templeton Foundation under Grant No. 63626.}

\bibliographystyle{eplbib}
\bibliography{0_references}

\end{document}